\documentclass[12pt]{article}

\usepackage{amssymb}
\usepackage{amsfonts}
\usepackage{amsmath}
\usepackage{amsthm}

\title{ Bild-conception for Scientific in Classical and Quantum Physics: from Hertz and Boltzmann to Schr\"odinger and De Broglie}

\author{ Andrei Khrennikov\\ 
Linnaeus University, International Center for Mathematical Modeling\\  in Physics and Cognitive Sciences
 V\"axj\"o, SE-351 95, Sweden\\
email: Andrei.Khrennikov@lnu.se}

\begin{document}
\maketitle

\begin{abstract}
We start with methodological analysis of the notion of scientific theory and its interrelation with reality.
This analysis is based on the works of Helmholtz, Hertz, Boltzmann, and Schr\"odinger (and reviews of D' Agostino). Following Helmholtz, Hertz established the ``Bild concept'' for scientific theories. Here 
``Bild'' (``picture'') carries the meaning ``model'' (mathematical). The main aim of natural sciences is construction of 
the causal theoretical models (CTMs) of natural phenomena. Hertz claimed that CTM cannot be designed solely on the basis of observational data; it typically contains hidden quantities. Experimental data can be described by an observational model (OM), often on the price of acausality. CTM-OM interrelation can be 
tricky. Schr\"odinger used the Bild concept to create CTM for quantum mechanics (QM) and QM was treated as OM. We follow him and suggest a special CTM for QM, so-called prequantum classical statistical field theory (PCSFT). QM can be considered as a PCSFT-image, but not as straightforward as in Bell's model with hidden variables. The common interpretation of the violation of the Bell inequality is criticized from the perspective of the two level structuring of scientific theories. Such critical analysis of von Neumann and 
Bell no-go theorems for hidden variables was performed already by De Broglie (and Lochak) in 1970s. 
The Bild-approach is applied to the two level CTM-OM modeling of Brownian motion: the overdamped regime corresponds to OM. In classical mechanics CTM=OM; on the one hand, this is very convenient, on the other hand, this exceptional coincidence blurred the general CTM-OM structuring of scientific theories. We briefly discuss ontic-epistemic structuring of scientific theories (Primas-Atmanspacher) and its relation to the Bild concept. Interestingly, Atmanspacher as well as Hertz claim that even classical physical theories should be presented on the basic of two level structuring.                
 \end{abstract}

{\bf keywords:} Bild conception; scientific theory;  Helmholtz; Hertz; Boltzmann; Schr\"odinger; De Broglie; quantum mechanics; Brownian motion; Bell, von Neumann

\section{Introduction}

The Bild-conception  of scientific theory was developed by Hertz \cite{Hertz,Hertz1} starting with Helmholtz analysis \cite{Helmholtz}  of interrelation between physical reality and scientific theory. This line of thinking was continued by Boltzmann \cite{Boltzmann,Boltzmann1} and in 1950's by Schr\"odinger \cite{SCHB}-\cite{SCHB6}. The articles of D' Agostino \cite{DAgostino0}-\cite{DAgostino} contain philosophically deep reviews on their works. The German word ``Bild'' is translated to English as ``picture''. But in relation to the analysis of the meaning of a scientific theory it has the meaning of a model, a mathematical model.   

Helmholtz pointed out that  a scientific theory does not describe reality as it is. A scientific theory structures our sensations and perceptions within a priori forms of intuition (cf. with Kant). Such structuring leads to models of reality reflecting some features of the 
environment of observers. Therefore the dream for creation of a ``true theory'' matching perfectly with natural phenomena is in contradiction with Helmholtz's philosophy of science. Observational data should be taken with caution. Helmholtz highlighted causality of the natural phenomena and for him the main task of a scientific theory is to reflect this causality. Thus, from his viewpoint the main aim of scientific  studies  is construction of 
the causal theoretical models (CTMs) of natural phenomena. Theoretical causality is an image of natural causality. In terms of cognition, causality of human reasoning reflects causality of natural processes and it was developed in during biological evolution, from the primitive forms of life to humans.  

Hertz followed Helmholtz' approach to scientific theory, but he claimed that generally CTM can't be designed solely on the basis of observational data and it typically contains hidden quantities.
(So, in physics hidden variables were employed long before the quantum revolution.) 
 Experimental data is described by an observational model (OM) which is often acausal. The CTM-OM interrelation can be 
tricky. This framework  Hertz  presented \cite{Hertz,Hertz1} as a Bild-conception  (model-concept). He  highlighted the role of mathematics and treatment of a scientific model as a mathematical model (see also Plotnitsky \cite{PL4}). In particular, Hertz presented 
Maxwell's theory as the system of the Maxwell equations.  

Later the Bild-conception  resurrected in the foundational studies of Schr\"odinger \cite{SCHB}-\cite{SCHB} (see especially \cite{SCHB}) who tried  to create CTM for quantum mechanics (QM) and QM was treated as OM. He advertized the two level structuring of the description of microphenomena.
 We follow him and suggest a special CTM for QM, so-called {\it prequantum classical statistical field theory} (PCSFT) \cite{Beyond,KHRB3,KHRAc}. QM treated as OM can be considered as a PCSFT-image, but not as straightforward as in Bell's model with hidden variables \cite{Bell0,Bell1}. 

We analyze Bell's model with hidden variables within the Bild-framework and criticize identification of subquantum (hidden) quantities with quantum observables and hidden probability distributions with quantum probability distributions. The evident barrier for such identification is the Heisenberg uncertainty principle (and the Bohr complementarity principle \cite{BR0,PL1,PL2}). The same viewpoint was presented long ago by  
De Broglie \cite{DeBroglie}  (see also Lochak \cite{Lochak,Lochak1,Lochak2}) who justified the legitimacy  of his double solution theory \cite{DB,BacciagaluppiV}, in fact, within the Bild-conception (although it seems that he was not aware about it). He pointed to inconsistency of the no-go interpretation of the von Neumann \cite{VN} and Bell \cite{Bell0,Bell1} theorems. De Broglie double solution model is an CTM for QM. (Its structuring within the Bild-conception deserves a separate article as well as the Bild-conception presentation of Bohmian mechanics.)\footnote{ This is the good place to note that one should not identify De Broglie's and Bohm's theories, not consider the later as just an extension and improvement of the former. De Broglie did not consider his theory as non-local, he noted that its nonlocality is only apparent, this is nonlocality of mathematical equations and not physical processes \cite{DB}. Deep foundational studies on De Broglie's double solution model are presented in works of Bacciagaluppi,e.g., \cite{Bacciagaluppi1,Bacciagaluppi2}, see also  appendixes written by Bacciagaluppi and Valentini in book \cite{BacciagaluppiV}.}    

We also use the Bild-approach for the two level CTM-OM modeling of Brownian motion: the overdamped regime corresponds to OM \cite{AKNJ}. Coarse gained velocities are observable quantities.  This example represents clearly the physical origin of the two level structuring of the mathematical description of Brownian motion. This is the time-scale separation technique. The evolution of the momenta of the
Brownian particles is very fast and cannot be resolved on the time-scales available 
to the experiment.  We notice that the OM model for the Brownian motion shows some distinguished properties of QM, see, e.g., article \cite{AKNJ}  for the corresponding uncertainty relations and Brownian entanglement theory.   

The idea of time-scale
separations is one of the most pertinent ones in non-equilibrium
statistical physics. In a qualitative form it appears already in good
textbooks on this subject
\cite{landau,ma}, and has been since then formalized in various
contexts and on various levels of generality
\cite{landauer,kurchan,theo,at,shu}.

In classical mechanics CTM=OM; on the one hand, this is very convenient, on the other hand, this exceptional coincidence blurred the general CTM-OM structuring of scientific theories.

We also briefly discuss ontic-epistemic structuring of scientific theories (Primas-Atmanspacher \cite{Primas,ATM}, see also artciles \cite{ATM1,ATM2}) and its relation to the Bild concept. 

This paper is continuation of my works \cite{KHRB3,KHRAc}. I hope that in this paper the Bild-conception and its implementation for quantum and classical mechanics  are presented 
clearer. Presentation  of the two level CTM-OM description for Brownian motion is a good complement to 
such description for quantum phenomena. The CTM-OM viewpoint on the Bell inequality project clarifies
the difference in the positions of Schr\"odinger \cite{SCHB}-\cite{SCHB6} and Bell \cite{Bell0,Bell1} on the possibility to construct a subquantum model with hidden variables.         

\section{Two level structuring of scientific theories}
\label{2level}

We start by citing the article fo D' Agostino \cite{DAgostino}:

\begin{footnotesize}
Hermann von Helmholtz (1821-1894) was one of the first scientists to criticize the
objective conception  of physical theory by denying that theoretical concepts describe
real physical objects. He realized that Immanuel Kant's a priori forms of intuition
should to be taken into account in analyzing problems that were emerging at the end
of the nineteenth century in the new formulations of physics. 

The objective conception  of physical theory also was criticized by such physicists as
Heinrich Hertz (1857-1894) and Ludwig Boltzmann (1844-1906), who adopted the
Kantian term Bild \footnote{Here the German word ``Bild'' (``picture'') is used in the sense of a model.} to designate the new conception  of physical theory, which they took
to mean not a faithful image of nature but an intellectual construct whose relationship
to empirical phenomena was to be analyzed.
\end{footnotesize}

The works of von Helmholtz, Hertz, and Boltzmann  \cite{Helmholtz,Hertz,Hertz1,Boltzmann,Boltzmann1} played the crucial role in development of a novel scientific methodology. Since the time of Galileo and Newton, scientific theories varied essentially in their content, but nobody questioned their ``ontological significance'', their adequacy to physical reality. 

In  1878, von Helmholtz posed the following philosophical questions \cite{Helmholtz}:

\begin{footnotesize}
What is true in our intuition and thought? In what sense do our representations correspond
to actuality?.
\end{footnotesize}

Von Helmholtz' answers to these questions were based on his physiological, especially visual, research that 
led him to the following conclusion \cite{Helmholtz}:  

\begin{footnotesize}
Inasmuch as the quality of our sensation gives us a report of what is peculiar to the
external influence by which it is excited, it may count as a symbol of it, but not as an
image. For from an image one requires some kind of alikeness with the object of
which it is an image ... .
\end{footnotesize}
We point out that if the fathers of QM would take this statement into account, then they surprise by ``unusual features'' of the quantum mechanical description of micro-phenomena would not be so strong. We also note that Bohr's views on QM match with this conclusion of Helmholtz. Surprisingly, it seems that Bohr had never referred to his works.  

Helmholtz's viewpoint on interrelation of sensations and generally observations and real objects led to the well known statement on the parallelism of the laws of nature and science \cite{Helmholtz}:

\begin{footnotesize}
Every law of nature asserts that upon preconditions alike in a certain respect, there
always follow consequences which are alike in a certain other respect. Since like
things are indicated in our world of sensations by like signs, an equally regular
sequence will also correspond in the domain of our sensations to the sequence of
like effects by the law of nature [that like effects follow from] ... like causes. 
\end{footnotesize}

We point out that the statement \begin{footnotesize} that upon preconditions alike in a certain respect, there
always follow consequences which are alike in a certain other respect \end{footnotesize} is about ontic causality.  So, for Helmholtz, nature is causal, i.e., laws in nature really exist and laws presented in scientific theories are mental representations of laws of nature. The laws expressed by our sensation and through them by our perception are ``parallel'' to natural laws, but only parallel, not identical, since our mind operates not with precise images of real objects, but only with symbols assigned to them.

Hertz questioned Helmholtz's parallelism of laws. Hertz believed that Helmholtz's parallelism of laws not only was indeterminate but in general even impossible if theory were limited to describing observable quantities \cite{Hertz1}: 

\begin{footnotesize}
If we try to understand the motions of bodies around us, and to refer them to simple
and clear rules, paying attention only to what can be directly observed, our
attempt will in general fail. We soon become aware that the totality of things visible
and tangible do not form an universe conformable to law, in which the same
results always follow from the same conditions. We become convinced that the manifold
of the actual universe must be greater than the manifold of the universe which
is directly revealed to us by our senses.
\end{footnotesize}

By Hertz a causal theory cannot be based solely on observable quantities \cite{Hertz1} : 
\begin{footnotesize} --- do not form an universe conformable to law, in which the same
results always follow from the same conditions.\end{footnotesize} Only by introducing hidden quantities Helmholtz's parallelism of laws can  become a general principle in physical theory. But such hidden
quantities (concepts that correspond to no perceptions) brings too much 
freedom in the choice of theoretical concepts.  To limit this freedom of choice, Hertz introduced special
requirements for the validation of a physical theory. Besides causality,  the most important was
theory's simplicity \cite{Hertz1}:

\begin{footnotesize}
It is true we cannot a priori demand from nature simplicity, nor can we judge what
in her opinion is simple. But with regard to images [Bilder] of our own creation we
can lay down requirements.We are justified in deciding that if our images are well
adapted to the things, the actual relations of the things must be represented by simple
relations between the images.
\end{footnotesize}

So, Helmholtz and Herz questioned the ontological status of scientific theories, as describing reality as it is. Scientific theories are only ``Bilder'', models of reality. Outputs of sensations and observations are just symbols encoding external phenomena. Hence, one should not sanctify observational quantities and their role in scientific theories. Moreover, an observational theory, i.e., operating with solely observables cannot be causal. Causality demands introduction of hidden (unobservable) quantities. Of course, a theory with hidden quantities should be coupled to observational data. However, this coupling need not be straightforward. 

According to Helmholtz a scientific theory should be causal. Hertz claimed \cite{Hertz1} that generally the 
causality  constraint requires invention of hidden quantities, a causal description  cannot be done solely  in terms of observational quantities. This approach unties scientists' hands, by introducing hidden quantities they can generate a variety of theoretical causal models coupled to the same observational quantities. How can one select a ``good'' causal model? Hertz suggested to use model's simplicity as a criterion for such selection. We note that even a ``good model'' does not describe reality as it is, it provides just a mathematical symbolic representation involving a variety of elements having no direct relation with observational quantities. 

It is natural to search for such (causal) theoretical model that would  describe what nature really is,  a ``true model'' (an ontic model). It is not clear whether Hertz might hope to design such a model 
for the electromagnetic phenomenon.\footnote{He tried to model it with systems of mechanical oscillators, i.e., to go beyond the electromagnetic field representation \cite{Hertz}. But he did not succeed with this project. His project was not meaningless. It has some degree of similarity with the representation of the quantum electromagnetic field as a system of quantum oscillators - photons.} Schr\"odinger who later contributed to development of the Bild concept of scientific theories, especially in the relation to the quantum foundations claimed \cite{SCHB1} that no true model can be formulated on the basis of our large-scale experience, because 

\begin{footnotesize}
we find nature behaving so entirely differently from what we observe in visible and
palpable bodies of our surroundings ... \;. A completely satisfactory model of this type
is not only practically inaccessible, but not even thinkable. Or, to be precise, we can,
of course, think it, but however we think it, it is wrong; not perhaps quite as meaningless
as a ``triangular circle,'' but much more so than a ``winged lion.''
\end{footnotesize}

Creation of a causal theoretical model coupled to some observed natural phenomena is a complex and long process. Moreover, there is always a chance that such a model would be never found - due to intellectual incapacity of humankind. Therefore it is natural to design models matching observations, but not satisfying the causality constraint. We call such models observational models. 

Thus, we distinguish two classes of models, {\it observational models} (OMs) and {\it causal theoretical models} (CTMs).  We remark that both kinds of scientific models are mental constructions, providing symbolic mathematical descriptions of natural phenomena. One may say that any model is theoretical, so OM is also theoretical. And he would be right. So, the main difference between OM and CTM is in causality. If OM is 
causal by itself, then there is no need to go beyond it with some CTM.    

Interrelation between CTM and OM, ${\bf M}_{\rm{CTM}}$ and ${\bf M}_{\rm{OM}},$  depends on the present stage of development of science. If ${\bf M}_{\rm{CTM}}$ rightly reflects the real physical processes, then development of measurement technology can lead to novel observational possibilities and some hidden quantities of ${\bf M}_{\rm{CTM}}$ can become measurable. Hence, ${\bf M}_{\rm{CTM}}$ becomes 
OM, ${\bf M}_{\rm{CTM}} \to {\bf M}_{\rm{OM}}^\prime.$ In principle, ${\bf M}_{\rm{OM}}^\prime$ need not 
cover all observations described by the previous OM ${\bf M}_{\rm{OM}}.$ New theoretical efforts might be needed to merge ${\bf M}_{\rm{OM}}$ and ${\bf M}_{\rm{OM}}^\prime.$ This abstract discussion will be illustrated by the concrete example from classical statistical physics - the two level modeling of the  Brownian motion (section \ref{SBrownian motion}).   

The ideas of Helmholtz and Hertz were further developed (and modified) in the works of Boltzmann 
\cite{Boltzmann,Boltzmann1}.  Then, 60 years later, Schr\"odinger \cite{SCHB}-\cite{SCHB6} contributed to development of the Bild viewpoint on quantum theories. He confronted with the special case of the aforementioned problem.

OM for micro-phenomena was developed (in particular, due to his own efforts): this is QM.   But QM suffered from acausality. The impossibility to solve the measurement problem (which was highlighted by von Neumann \cite{VN}) generates a gap in the quantum description of micro-phenomena. Schr\"odinger came back to this problem in 1950's \cite{SCHB}- \cite{SCHB6}; this comeback was stimulated by development of quantum field theory and the method of second quantization. 

He saw in quantum field theory a possibility to justify his attempts of the purely wave (continuous) approach to modeling of the micro-phenomena. In complete agreement with the Bild concept,  he considered QM as an observational model. As well as von Neumann,  Schr\"odinger  highlighted its acausality. But it was not treated as a property of nature as it is, i.e., quantum acausality (of measurements and spontaneous quantum events) is not ontic. We notice that, for von Neumann, it is ontic, he wrote about ``irreducible quantum randomness'' \cite{VN}. Quantum acausality is just a property of special OM - QM. Schr\"odinger claimed that quantum acausality is related to ignoring of the Bild concept and assigning the ontological status to quantum particles, see his article ``What is an elementary particle?'' \cite{SCHB}. We remark that Bohr did not question the ontological status of quantum systems, atoms, electrons and may be even photons \cite{BR0,PL1,PL2}.  Schr\"odinger considered indistinguishability of quantum particles as a sign that they do not have the ontological status. Hence, instead of OM (= QM), one can hope to develop CTM for microphenomena, by liberating it from particles and operating solely with waves. 

Since waves propagate in space, for Schr\"odinger causality (in fact, the wave causality) is coupled to continuity in the space, so the waves should be continuous (see Plotnitsky \cite{PL4} on analysis of continuity vs. discontinuity in physics). We remark that he considered continuity of waves on  multi-dimensional space $\mathbb{R}^{3n}.$ In 1920s the fact that the multi-particle Schr\"odinger equation describes the waves not on ``the physical space'' $\mathbb{R}^3,$ but on ``the mathematical space'' 
$\mathbb{R}^{3n},$ was disturbing for him. This was the main reason for Schr\"odinger  to accept the probabilistic interpretation of the wave function. At that time he did not use the Bild concept for
scientific theories (was not aware about the works of Helmholtz, Herz, and Boltzmann?). By the Bild concept 
the wave representation of QM is just a symbolic mathematical representation of the micro-phenomena. The use 
of the multi-dimensional space $\mathbb{R}^{3n}$ has the same descriptive status as the use of $\mathbb{R}^3.$

Schr\"odinger dreamed for creation of CTM for micro-phenomena, his concrete intention was towards a wave-type model. He also highlighted the principle fo continuity for ``quantum waves'', but he suspected that it would be valid only at the microlevel. He pointed to quantum field theory 
as a good candidate to proceed in this direction. Since he coupled causality and continuity, it became possible to relax the causality-continuity constraint  and restrict this constraint to the level of infinitesimals. In a theoretical model completing QM (an observational model) for which Schr\"odinger dreamed, causality need not be global. 

Schr\"odinger's continuous wave completion project for QM has some degree of similarity with Einstein's project on designing a classical field model of micro-phenomena which he announced with Infeld in a popular form in book \cite{Infeld}.\footnote{Einstein's intention was that a complete theory beyond QM 
should be non-linear field theory. Later Infeld contributed a lot into this project. In contrast tyo Einstein, Schr\"odinger dreamed for a linear model.}  However, in contrast to Schr\"odinger, Einstein did not appeal to the Bild concept on the  two level modeling of natural phenomena, observational and causal theoretical (OM and CTM), and a possible gap between these two models. The presence of such gap, in particular, implies that CTM need not describe the observational data straightforwardly.  

Einstein's project on reconsideration of quantum foundations starting with the EPR-paper \cite{EPR} was not directed to the two level structuring of the mathematical description of microphenomena. He dreamed for CTM which would match perfectly with quantum observations. This dream was later formalized by Bell in his hidden variables model \cite{Bell0,Bell1}.

Schr\"odinger understood \cite{SCHB1} that  CTM of microphenomena of the wave type  is not
 \begin{footnotesize}
the observed or observable facts; and still less do we claim that we thus describe what nature (matter, radiation, etc.) really is. In fact we use this picture (the so-called wave picture) in full
knowledge that it is neither.   
\end{footnotesize} 
This statement expresses the extreme view on the Bild concept;  Schr\"odinger \cite{SCHB1} also pointed out that 
 \begin{footnotesize} observed facts ... appear to be repugnant to the classical
ideal of a continuous description in space and time.\end{footnotesize} Such highlighting of decoupling 
of theory and observations was too provocative and played the negative role. The idea of using the Bild concept in quantum foundations was rejected by the majority of experts in quantum foundations.

However, the Bild concept did not disappear completely and its trace can be found in the philosophy of 
the ontic-epistemic  structuring of physical theories that was developed by 
Primas and Atmanspacher \cite{ATM} (see also, e.g.,  \cite{ATM1,ATM2}). 
They tried to find an answer  \cite{ATM1} to the old question: \begin{footnotesize}
Can nature be observed and described as it is in itself independent of those
who observe and describe - that is to say, nature as it is ``when nobody
looks''? 
\end{footnotesize}

As well as Helmholtz, Hertz, Boltzmann, and Schr\"odinger, they pointed out that observations give to observers only some knowledge about systems, this knowledge is incomplete. This knowledge is mathematically structured within an epistemic (=observational) model . For them, QM is such a model, i.e., w.r.t. QM the views of Schr\"odinger and  Primas-Atmanspacher coincide. Then, in the same way as Schr\"odinger, they want to have a complete model of microphenomena. The crucial difference from the Bild concept is that  Primas and Atmanspacher were seeking for an ontic model, a model of reality as it is, the ``true model'' in terms of Schr\"odinger. Generally Primas and Atmanspacher also supported the idea of the two level structure of scientific theories: epistemic (observational) and ontic. As well as Schr\"odinger, they pointed out that the connection between  epistemic and ontic models is not straightforward. Causality is the basic property of the ontic model.  So, if one would ignore the term ``ontic''
\footnote{Its use would be very disturbing for Helmholtz, Hertz, Boltzmann, and Schr\"odinger.}, then formally (and mathematically)  Primas-Atmanspacher structuring of the scientific description of nature is  similar to the Bild concept. (In contrast to Schr\"odinger, they did not emphasize the continuous wave structure of an ontic model beyond QM.) 

However, by pointing to formal mathematical similarity of the ontic-epistemic and Bild approaches, one should remember that they differ crucially from the foundational perspective.   We recall \cite{ATM1} that
\begin{footnotesize} 
Ontological questions refer to the structure and behavior of a system as
such, whereas epistemological questions refer to the knowledge of information
gathering and using systems, such as human beings.
\end{footnotesize}

From the Bild perspective, it is totally meaningless even to \begin{footnotesize} refer to the structure and behavior of a system as such ... \end{footnotesize}  The essence of the ontic-epistemic approach is expressed in the following quote from Atmanspacher \cite{ATM1} (for more details, the reader is referred to Primas \cite{Primas} ):

\begin{footnotesize}
Ontic states describe all properties of a physical system exhaustively. (``Exhaustive'' in this context means that an ontic state
is ``precisely the way it is'', without any reference to epistemic
knowledge or ignorance.) Ontic states are the referents of 
individual descriptions, the properties of the system are treated as
intrinsic properties.Their temporal evolution (dynamics) is reversible 
and follows universal, deterministic laws. As a rule, ontic
states in this sense are empirically inaccessible. Epistemic states
describe our (usually non-exhaustive) knowledge of the proper-
ties of a physical system, i.e. based on a finite partition of the
relevant phase space. The referents of statistical descriptions are
epistemic states, the properties of the system are treated as con-
textual properties. Their temporal evolution (dynamics) typically
follows phenomenological, irreversible laws. Epistemic states are,
at least in principle, empirically accessible
\end{footnotesize}

From the Bild perspective, the statement: \begin{footnotesize} Ontic states are the referents of 
individual descriptions, the properties of the system are treated as
intrinsic properties,\end{footnotesize} is meaningless, since systems do not have intrinsic properties,
a theoretical causal model beyond the quantum observational (epistemic) model still describes not the properties of  the systems, but our mental pictures.

And we conclude this section by the quote from Nietzsche (written in 1873, but published later); his statement is very similar  similar to Helmholtz's statements, but it is more passionate or even poetic!  It seems that Nietzsche was influenced by Helmholtz, especially on nerve stimulus.  Nietzsche wrote about language, but the point is more general \cite{Nietzsche}\footnote{I would like to thank Arkady Plotnitsky for mentioning this quote in our discussion on the works of Helmholtz, Hertz, Boltzmann, and Schr\"odinger and especially for sending to me this reference to Nietzsche.}:

\begin{footnotesize}
The various languages placed side by side show that with words it is never a question of truth, never a question of adequate expression; otherwise, there would not be so many languages. The ``thing in itself'' (which is precisely what the pure truth, apart from any of its consequences, would be) is likewise something quite incomprehensible to the creator of language and something not in the least worth striving for. This creator only designates the relations of things to men, and for expressing these relations he lays hold of the boldest metaphors. To begin with, a nerve stimulus is transferred into an image: first metaphor. The image, in turn, is imitated in a sound: second metaphor. And each time there is a complete overleaping of one sphere, right into the middle of an entirely new and different one. One can imagine a man who is totally deaf and has never had a sensation of sound and music. Perhaps such a person will gaze with astonishment at Chladni's sound figures; perhaps he will discover their causes in the vibrations of the string and will now swear that he must know what men mean by ``sound.'' It is this way with all of us concerning language; we believe that we know something about the things themselves when we speak of trees, colors, snow, and flowers; and yet we possess nothing but metaphors for things-metaphors which correspond in no way to the original entities. In the same way that the sound appears as a sand figure, so the mysterious of the thing in itself first appears as a nerve stimulus, then as an image, and  finally as a sound. Thus the genesis of language does not proceed logically in any case, and all the material within and with which the man of truth, the scientist, and the philosopher later work and build, if not derived from never-never land, is a least not derived from the essence of things. 
\end{footnotesize}

\section{Coupling of theoretical and observational models}
\label{CC}

Models considered in natural science are mainly mathematical. Therefore coupling between CTM and OM corresponding to the same natural phenomena is a mapping of one mathematical structure to another.

Consider some mathematical model ${\bf M},$ either CTM or OM. It is typically based on two spaces, 
the space of states $S$ and the space of variables (quantities) $V.$ For OM, $V$ is the space of observables, instead of states one can consider measurement contexts.  

Consider  OM model ${\bf M}_{OM}$ and its causal theoretical completion  ${\bf M}_{CTM}.$  It is natural to have a mathematical  rule establishing  correspondence between them.   We recall that CTMs are causal and OMs are often acausal; if it happens that OM is causal, then there is no need for a finer description given by some CTM. Thus, the task is to establish correspondence between causal and acausal models. It is clear that such correspondence cannot be straightforward. We cannot map directly states from $S_{CTM}$ to states
from $S_{OM}.$   Causality can be transformed into acausality through consideration of probability distributions.  So, consider some space  of probability distributions $P_{CTM}$ 
on the state space $S_{CTM}$ and construct a map from $P_{CTM}$ to $S_{OM},$ the state space of OM.
This approach immediately implies that  the states  of OM are interpreted statistically.   We also should establish  correspondence between variables (quantities) of ${\bf M}_{CTM}$ and ${\bf M}_{OM}.$ Thus, we 
need to define two physically natural maps: 
\begin{equation}
\label{m1}
J_S:  P_{CTM} \to  S_{OM}, \; J_V:  V_{CTM} \to  V_{OM}.
\end{equation}
Since  $J_S$ is not defined for states of CTM, but only for probability distributions, ``physically natural'' means coupling between the  probability structures of ${\bf M}_{CTM}$ and ${\bf M}_{OM};$
the minimal coupling is the equality of averages
between variables 
\begin{equation}
\label{cor}
\langle J_V(f) \rangle_{J_S(P)} = \langle f \rangle_P
\end{equation}
and correlations 
\begin{equation}
\label{cor1}
 \langle J_V(f) J_V(g)\rangle_{J_S(P)}= \langle f g \rangle_P.
\end{equation}
Generally the correlation need not be defined, so (\ref{cor1}) should hold for variables $f, g \in V_{CTM}$ and observables $A_f=J_V(f)$ and $A_g=J_V(g)$ for which the correlations in the states  
$P$ and $J_S(P)$ are defined.

Mathematically causality can be realized as functional representation of variables (see  monograph of Wagner
\cite{Wagner} on such representation of causality). Therefore we assume that $V_{CTM}$ can be represented as a space of functions $f: S_{CTM} \to \mathbb{R}.$ Such model is causal, the state $\phi$ uniquely determines the values of all  variables belonging $V_{CTM}: \phi \to f(\phi).$  
The state space $S_{CTM}$ can be  endowed with a $\sigma$-algebra of subsets ${\cal F}.$ Elements of 
 $P_{CTM}$ are probability measures on ${\cal F}.$ The minimal mathematical restriction on elements of $V_{CTM}$ is that they are measurable functions, $f: S_{CTM} \to \mathbb{R}.$ In such a framework,
\begin{equation}
\label{cor2}
\langle f  \rangle_P= \int_{ S_{CTM}} f(\lambda) P(d\lambda),
\langle f g \rangle_P= \int_{ S_{CTM}} f(\lambda) g(\lambda) P(d\lambda),
\end{equation}
if the integrals exist, e.g., if CTM-variables are square integrable: 
$$
\int_{ S_{CTM}} |f(\lambda)|^2  P(d\lambda) < \infty.
$$

Since in ${\bf M}_{OM}$ quantities have the experimental statistical verification, we establish some degree of experimental verification for ${\bf M}_{CTM}$ through mapping of ${\bf M}_{CTM}$ to ${\bf M}_{OM}.$   But such verification is only indirect, one should not expect direct coupling between quantities of ${\bf M}_{CTM}$ and experiment (as Einstein, Bell and all their followers wanted to get).  
Generally these maps are neither one-to-one nor onto. 
\begin{itemize}
\item A cluster of probability distributions on $S_{CTM}$  can be mapped into the same  state from $S_{OM}.$
\item  $J_S(P_{CTM})$ need not coincide with $S_{OM}.$ 
 \item A cluster of elements of $V_{CTM}$ can be mapped into a single variable (observable) from
 $V_{OM}.$
\item $J_V(V_{CTM})$ need not coincide with $V_{OM}.$ 
\end{itemize}

Moreover, the model-correspondence maps $J_S, J_V$ need not be defined on whole spaces 
$P_{CTM}$ and $V_{CTM}.$ They have their domains of definition, ${\cal D}_{J_S} \subset P_{CTM}$ 
and ${\cal D}_{J_V} \subset V_{CTM}.$ (In principle, one can reduce $P_{CTM}$ to
$P_{CTM}^\prime = {\cal D}_{J_S}$ and $V_{CTM}$ to $V_{CTM}^\prime={\cal D}_{J_V}$ and operate with  
maps $J_S, J_V$ which are defined everywhere on these reduced spaces of CTM's states and variables).

We remark that the same ${\bf M}_{OM}$ can be coupled to a variety of CTMs. We also remark that the same 
observational data can be mathematically described by a variety of OMs. 

We also remark that similarly to the deformation quantization (here we discuss just the mathematical similarity) CTM may depend on some small parameter $\kappa$ (in the deformation quantization this is action, roughly speaking the Planck constant $h).$ Thus, ${\bf M}_{CTM}= {\bf M}_{CTM}(\kappa).$ In such more general framework, the correspondence maps also depend on $\kappa,$ i.e.,  
$J_S= J_S(\kappa), J_V= J_V(\kappa).$ The probabilistic coupling constraints (\ref{cor}), (\ref{cor1}) can be weakened:
\begin{equation}
\label{cora}
\langle J_V(\kappa;f) \rangle_{J_S(\kappa; P)} = \langle f \rangle_P + o(\kappa), \kappa\to 0,  
\end{equation}
\begin{equation}
\label{cora1}
 \langle J_V(\kappa; f) J_V(\kappa; g)\rangle_{J_S(\kappa;P)}= \langle f g \rangle_P  +o(\kappa), \kappa\to 0 
\end{equation}
(see \cite{}). The problem of identification of the parameter $\kappa$ with some physical scale is complex
(see, e.g., \cite{KHRB1,KHRB2} for an attempt of such identification within PCSFT).     

\section{Prequantum classical statistical field theory as a causal theoretical model for quantum mechanics}
\label{PCFTBild}
 
We illustrate the general scheme of CTM-OM correspondence by two theories of micro-phenomena,  QM as ${\bf M}_{OM}$ and  PCSFT  as ${\bf M}_{CTM}.$ Re-denote these model with the symbols  ${\bf M}_{\rm{QM}}$ and 
${\bf M}_{\rm{PCSFT}}.$  We briefly recall the basic elements of PCSFT (see \cite{Beyond,KHRB3,KHRAc}  for details).

In ${\bf M}_{QM}$  states are given by density operators acting in complex Hilbert space ${\cal H}$ (with scalar product $\langle \cdot\vert  \cdot \rangle)$ 
and observables  are represented by Hermitian operators in ${\cal H}.$ Denote the space of density operators by $S_{\rm{QM}}$ and the space of Hermitian operators by $V_{\rm{QM}}.$ 

In ${\bf M}_{\rm{PCSFT}}$ states are vectors of ${\cal H},$ i.e., 
$S_{\rm{PCSFT}} ={\cal H}.$   Physical variables  are quadratic forms
$$
\phi \to f(\phi) = \langle \phi \vert  \hat  A \vert  \phi \rangle,
$$ 
where $ \hat  A\equiv  \hat A_f$ is a Hermitian operator. The space of quadratic forms is denoted by the symbol 
$V_{\rm{PCSFT}}.$ Consider probability measures on the $\sigma$-algebra of Borel subsets  of ${\cal H}$ 
(i.e., generated by balls in this space)
having  zero first momentum, i.e., 
\begin{equation}
\label{m3}
\int_{{\cal H}} \langle \phi \vert a\rangle d p(\phi)=0
\end{equation}
for any vestor $a\in  H,$ and finite second momentum, i.e., 
\begin{equation}
\label{m4}
{\cal E}_p \equiv \int_{{\cal H}} \Vert \phi \Vert^2 d p(\phi) < \infty.
  \end{equation}
	Denote the space of such probability measures by the symbol $P_{\rm{PCSFT}}.$ 
	
We can start not with probability measures, but with  ${\cal H}$-valued random vectors with zero mean value and finite second moment: $\phi= \phi(\omega),$  such that  $E [\phi]= 0$ and $E[\Vert \phi\Vert^2] < \infty.$\footnote{Random vectors are defined on some Kolmogorov probability space $(\Omega, {\cal F}, P),$ these are functions $\phi: \Omega \to {\cal H}$ which are measurable w.r.t. to the Borel $\sigma$-algebra of ${\cal H},$ i.e., for any Borel subset $B$ of ${\cal H},$ $\phi^{-1}(B) \in {\cal F}.$ A map is measurable iff, for any $c>0,$ the set $\Omega_{\phi,c}= \{\omega \in \Omega: || \phi(\omega)|| < c\} \in {\cal F}.$}
The space of such random vectors is denoted by the symbol $R_{\rm{PCSFT}}.$
In the finite-dimensional case, these are complex vector-valued random variables; if ${\cal H}$ is 
infinite-dimensional, then the elements of  $R_{\rm{PCSFT}}$ are random fields.

An example of random fields is given by selection ${\cal H}=L_2(\mathbb{R}^n; \mathbb{C})$ of square integrable complex valued functions. Each ${\bf M}_{CTM}$ state $\phi$ is an $L_2$-function, $\phi: \mathbb{R}^n  \mapsto \mathbb{C}.$ 
Random fields belonging to  $R_{\rm{PCSFT}}$ are functions of two variables, $\phi= \phi(x; \omega):$ chance parameter $\omega$ and space coordinates 
$x.$  

We remark that, for the state space ${\cal H}=L_2(\mathbb{R}^n; \mathbb{C}),$ the quantity ${\cal E}_p$ can be represented as
$$
{\cal E}_p = \int_{{\cal H}}  {\cal E}(\phi) d p(\phi),
$$
where 
$$
{\cal E}(\phi) = \Vert \phi\Vert^2 = \int_{\mathbb{R}^n} \vert\phi (x) \vert^2 dx
$$
is the energy of the field. The quantity ${\cal E}_p$ can be interpreted as the average of the field energy with respect to the probability distribution $p$ on the space of fields. We can also use the 
random field representation. Let $\phi= \phi(x; \omega)$ be a random field. Then its energy is the random variable 
$$
{\cal E}_\phi(\omega)= \int_{\mathbb{R}^n} \vert\phi (x; \omega) \vert^2 dx
$$ 
and ${\cal E}_p$ is its average. 

For any $p \in P_{\rm{PCSFT}},$ its (complex) covariance operator $\hat B_p$ is defined by its bilinear (Hermitian) form:
\begin{equation}
\label{m5}
\langle a \vert  \hat  B_p\vert b\rangle = \int_{{\cal H}} \; \langle a\vert \phi  \rangle  \langle \phi \vert b  \rangle \; d p(\phi), \; a, b \in {\cal H}, 
\end{equation}
or,  for a random field $\phi,$ we have:   
$$
\langle a \vert \hat B_\phi\vert b\rangle = E  [\langle a\vert \phi \rangle  \langle \phi \vert b  \rangle].
$$
We note that 
\begin{equation}
\label{m4a}
{\cal E}_p  = \int_{{\cal H}} \Vert \phi \Vert^2 d p(\phi) = \rm{Tr} \hat B_p
 \end{equation}
or in terms of a random field:
\begin{equation}
\label{m4aa}
{\cal E}_p = E[||\phi||^2]= 
E[\int_{\mathbb{R}^n} \vert \phi (x; \omega) \vert^2 dx] =  \rm{Tr} \hat B_p.
 \end{equation}
Thus, the average energy of a random field $\phi=\phi(\omega, x)$ can be expressed via its covariance operator. 

Generally a probability measure (${\cal H}$-valued random variable ) is not determined by its covariance operator (even under the constraint given by zero average).

A complex covariance operator has the same  properties as a density operator, besides normalization by the trace one; a covariance operator $\hat B_p$ is 
\begin{itemize}
\item Hermitian,
\item positively semidefinite,
\item trace class.
\end{itemize}

A ``physically natural coupling'' of the models ${\bf M}_{\rm{QM}}$ and ${\bf M}_{\rm{PCSFT}}$  is based on   the following formula coupling mathematically the averages for these models. 
For a probability measure $p \in P_{\rm{PCSFT}}$ and a variable $f \in V_{\rm{PCSFT}},$ we have
\begin{equation}
\label{m6}
\langle f \rangle_p = \int_{{\cal H}} f(\phi) d p(\phi)= \rm{Tr} \hat A_f \hat B_p,
\end{equation}
where $f(\phi) = \langle \phi\vert \hat A_f\vert \phi\rangle.$ 
This formula is obtained through expansion of the quadratic form 
$\langle \phi\vert \hat A_f\vert \phi\rangle$ w.r.t. the basis of eigenvectors of the Hermitian operator 
$\hat A_f.$   
 
Let us consider the following maps $J_S: P_{\rm{PCSFT}} \to S_{\rm{QM}}$ and $J_V:V_{\rm{PCSFT}} \to V_{\rm{QM}}$, 
 \begin{equation}
\label{m1q}
J_S(p)=  \hat  \rho_p= \hat B_p/\rm{Tr} B_p, \; J_V(f)=   \hat A_f .
\end{equation}
This correspondence connects the averages given by the causal theoretical and observational models:  
\begin{equation}
\label{m6q}
\frac{1}{{\cal E}_p}\langle f \rangle_p  = \rm{Tr}  \hat  \rho_p  \hat  A_f ,
\end{equation}
i.e., the QM and PCSFT averages are coupled with the scaling factor which is equal to the inverse of the average energy of the random field (for ${\cal H}= L_2).$

Thus, density operators representing quantum states correspond to covariance operators of random fields  normalized by the average energy of a random field and the Hermitian operators representing 
quantum observables  correspond to quadratic forms of fields. 

Let us  rewrite (\ref{m6q}) in the form: 
$$
\langle \frac{f}{{\cal E}_p} \rangle_p  = \rm{Tr} \hat \rho_p \hat A_f.
$$ 
If random fields have low energy, i.e., ${\cal E}_p<<1,$ 
the quantity $$
g_p(\phi)\equiv \frac{f(\phi)}{{\cal E}_p}
$$ can be interpreted as an amplification of 
 the PCSFT physical variable $f.$ Hence,  by connecting QM with  PCSFT,  QM can be interpreted as an observational theory describing averages of amplified
`subquantum' physical variables - quadratic forms of random fields. The subquantum random fields are unobservable and they can be experimental verified only indirectly, via coupling with the observational 
model - QM. 

In contrast to QM, PCSFT is causal: selection of a vector (`field') $\phi \in {\cal H}$ determines the values of all PCSFT-variables, quadratic forms  of classical fields:
$\phi \to  \langle \phi\vert \hat A\vert \phi\rangle.$ 

For physical variables, the correspondence  map $J_V$   is one-to-one, but  the map $J_S$ is not one-to-one. But it is a surjecion, i.e., it is on-to map.

\section{On usefulness of causal theoretical models} 
\label{use}

The above presentation of the possible two level description of the microphenomena, QM vs. PCSFT, can be used as the initial point for the discussion on usefulness of CTMs. To be provocative, we start by noting that for Bohr and other fellows of the orthodox Copenhagen interpretation of QM attempts to construct CTM for QM is meaningless \cite{BR,BR0,PL1,PL2}. At the same time Bohr never claimed that such CTM can't be constructed \cite{PL1,PL2}; he was not interested in no-go theorems. In his writings I did not find any word about the von Neumann no-go theorem. I 
am sure that he would ignore the Bell no-go theorem \cite{Bell1} and be surprised by interest to it in the modern quantum foundational community. Bohr highlighted the observational status of QM, but for him  any kind of CTM is a metaphysical. For ``real physicist'', it is meaningless to spend time by trying to design prequantum CTM. This position is very common among ``real physicists''.\footnote{ 
At the beginning of this century academician Kamil Akhmetovich Valiev (one of the engines of the quantum computer project in Russia) told me: ``Andrei, why do you consume your days for such totally meaningless activity? You are talented and can contribute so much to real physics - quantum computing!'' He pointed to my attempts to find a prequantum CTM.} For Bohr, it is impossible to complete QM in the causal way by operating with quantities which have direct connection to observations. And he is completely right: the complementarity principle and Heisenberg uncertainty relation block searching of a finer OM for QM. 
It seems that in contrast to von Neumann, Bohr was not disturbed by acausality, observational acausality is a consequence of contextuality and complementarity of quantum observations. We also repeat that Bohr did not deny the possibility to construct CTMs beyond QM, but for him the introduction of hidden variables was a metaphysical and totally meaningless exercise.   

Bohr's position can be questioned and on questioners' side are Helmholtz, Herz, Boltzmann, Schr\"odinger. As we have seen, Schr\"odinger agreed with Bohr that QM is a good OM for microphenomena. He did not think that acausal structure of QM prevents 
construction of corresponding CTM. For him causality is closely coupled to continuity and hence to his original wave approach to microphenomena.

As   Helmholtz, Herz, Boltzmann, Schr\"odinger, I think that consideration of acusality as the property of nature (at least at the micro-level) destroys completely the  methodology of science. If Helmholtz did mistake by saying \cite{Helmholtz} that

\begin{footnotesize} 
every law of nature asserts that upon preconditions alike in a certain respect, there
always follow consequences which are alike in a certain other respect,
\end{footnotesize}  
then physics becomes a science about gambling (as, e.g., QBists claim). It is difficult (at least for me)
to accept this position.  Thus, the main impact of creation of CTMs is reestablishing of causality that might be violated in OM. 

Now we turn to QM. Reestablishing of causality of microphenomena (even without the direct coupling to observations) would demystify quantum theory. We do not claim that PCSFT is the ``true CTM'' for QM; as 
Schr\"odinger claimed \cite{SCHB1} it is meaningless and even dangerous for science development to search for such a model.
But PCSFT can be used as a causal Bild of quantum processes. One of the main advantages of this Bild is that 
it is local. PCSFT reproduces not only QM averages, but event its correlations \cite{Beyond}; hence, the Bell inequalities can be violated for PCSFT-variables (hidden variables from the observational viewpoint). 
PCSFT demystifies quantum entanglement by connecting it with correlations of subquantum (classical) random fields (cf. \cite{Beyond,KHRB3,KHRAc}). PCSFT can be considered as a step towards merging of QM with general relativity, but within some CTM.

Can one earn from CTM something that might be lifted to the observational level? In our concrete case, can some theoretical elements of PCSFT be realized experimentally (may be in future)? The basic element of PCSFT is a random field $\phi= \phi(x; \omega).$ Measurement of such a subquantum field would be the real success of PCFT. However, it seems that one cannot expect this. As was pointed out by Bohr (in 1930s), even the classical electromagnetic field cannot be measured in a fixed point.  

Another component of PCSFT which can be connected with real physics is the need in the background field. This component was not discussed in the above brief presentation, so see \cite{Beyond,KHRBE} for details. Such random background field $\phi_{\rm{bground}}(x, \omega)$ is the necessary element of the mathematical model ${\bf M}_{\rm{PCSFT}}$  for generation of entangled states in $S_{\rm{QM}}.$ In this way PCSFT is related to stochastic  electrodynamics and supports it. Unfortunately, the background (zero point field) is not a component of conventional QM;  stochastic  electrodynamics  is commonly considered as unconventional model of microphenomena. From the Bild-viewpoint this model should be treated as one of possible CTM for QM, this viewpoint would clarify the interrelation between these two models. But in this paper we do not plan to go deeper in this issue. We note the background field carries long-distance correlations which contribute into violation of the Bell type inequalities. However, these are classical field correlations having nothing to do with the spooky action at a distance.    

The most close to experimental verification is PCSFT representation of Born's rule as an approximate rule for calculation of probabilities, the standard Born rule is perturbed by additional terms which can be in principle verified (see \cite{Beyond} and especially article \cite{KHRBE} suggesting the concrete experimental test). 
By totally different reason  this prediction was tested by the research group
of prof. Weihs \cite{Weihs,Weihs1}   -  testing Sorkin's inequality in the triple slit experiment. Surprisingly, transition from two slits to three slits is not trivial and Weih's group confronted difficulties related to nonlinearity of detection processes. For the moment, no deviations from the Born rule were observed.

For me, the main message of PCSFT as one of possible CTMs for QM is that ``quantum nonlocality'' is an artifact of OM (=QM). The presence of such artifacts in OMs is natural from the Bild-viewpoint. This is one of the reasons to construct CYMs.   
   
\section{Bell's project from the Bild-viewpoint}
\label{BellBild}

Unfortunately, Bell (as well as other ikons of modern quantum foundational studies)  did not read the works on the Bild concept for scientific theories. By introducing hidden variables he suggested some CTM for 
OM=QM. However, he considered too naive coupling of his CTM,  ${\bf M}_{\rm{Bell}},$ with OM -
${\bf M}_{\rm{QM}}.$ The subquantum quantities, functions of hidden variables, $A=A(\lambda),$  were 
identified with quantum observables. In particular the ranges of values of quantities from
${\bf M}_{\rm{Bell}}$ coincide with the ranges of values of quantum observables (and this is not the case e.g. in PCSFT). As was pointed out in article \cite{NL1},  ${\bf M}_{\rm{Bell}},$ confronts 
the complementarity principle. The later point will be clarified below. 

We recall the mathematical structure of ${\bf M}_{\rm{Bell}}$ by connecting it with the framework of section 
\ref{CC}. Bell considered \cite{Bell1} an arbitrary set of hidden variables $\Lambda;$ this is the set of states 
of his CTM, i.e., $S_{Bell}= \Lambda.$ To put this model in the mathematical framework of probability theory,  $\Lambda$ should be endowed with some $\sigma$-algebra of its subsets, say ${\cal F}.$ Denote by $P_{Bell}$ the space of all probability measures on $(\Lambda, {\cal F}).$ The space of Bell-variables consists of all measurable functions $A: \Lambda \to \mathbb{R}, A=A(\lambda),$ i.e., random variables in terms of  
the Kolmogorov probability theory. We stress that the correspondence map $J_S: P_{Bell} \to S_{QM}$ is not specified, 
it is just assumed that such map exists. For Bell's reasoning \cite{Bell1}, this map need not be onto $S_{QM}$ (so it need not be a surjection). 

To make his model with hidden variables straightforwardly experimentally verifiable, Bell identified the values of CTM quantities $A=A(\lambda)$ with the values of QM quantities, the outcomes of quantum observables. First we discuss the mathematical side of this assumption and then its foundational side.   

Mathematically identification of quantities of  ${\bf M}_{\rm{Bell}}$ with QM observables
means that the range of values of $A \in V_{Bell}$ coincides with the spectrum of the corresponding Hermitian operator $\hat A.$ This is the important mathematical constraint on the map $J_V :  V_{Bell}\to V_{QM}$ (we recall that $V_{QM}$ is the set of density operators). Purely mathematical relaxation of this assumption destroys the Bell inequality argument, e.g., as in PCSFT. 

However, Bell {\it should} proceed with this assumption on the coincidence of ranges of values of the subquantum quantities and quantum observables, since he dreamed for straightforward experimental verification of his model with hidden variables \cite{Bell0,Bell1}. He was not accustomed with the Bild concept of a scientific theory. In particular, Herz's (and Schr\"odinger's) statement on hidden quantities which could not be observed directly was totally foreign for Bell. For him, as well as for Bohr, a theory which quantities can't be directly verified is a part of metaphysics, not physics \cite{BR,BR0,PL1,PL2}.

 However, by identifying the outcomes of subquantum quantities with the outcomes of quantum observables Bell confronts the complementarity  principle. This can be clearly seen in the CHSH-framework \cite{CHSH}. There are two pairs of observables: $(A_1, A_2),$ in ``Alice's lab'', and $(B_1,B_2),$ in  ``Bob's lab'' represented by Hermitian operators $(\hat A_1, \hat A_2)$  and $(\hat B_1, \hat B_2).$
Observables corresponding to cross measurements for Alice-Bob are compatible, i.e., they can be jointly measurable, but local observables of Alice as well as of Bob are incompatible, i.e., they cannot be jointly measurable. In the operator terms, 
\begin{equation}
[\hat A_i, \hat B_j]=0, \;  [\hat A_1, \hat A_2]\not=0, \; [\hat B_1, \hat B_2] \not=0.
\end{equation}
This is the quantum mechanical description of the CHSH experimental context. We note that if local observables are compatible for at least one lab (in the operator terms at least one of commutators 
$[\hat A_1, \hat A_2], [\hat B_1, \hat B_2]$ equals to zero), then the CHSH inequality can't be violated \cite{NL1}. Bell considered variables of his CTM ${\bf M}_{\rm{Bell}}$ as representing physical observables, hence all observables can be represented as functions $A_i=A_i(\lambda), B_j=A_i(\lambda)$ and their values are identified with outcomes of observations. Besides the pairs $(A_i(\lambda),B_j(\lambda))$ of compatible 
observables, one can consider the pairs $(A_i(\lambda),A_j(\lambda))$ and $(B_i(\lambda),B_j(\lambda)).$
By treating the later two pairs as representing the outcomes of physical observables it is natural to assume the possibility of their joint measurability, may be not nowadays, but in future, when the measurement technologies would be improved. So, the complementarity principle loses its fundamental value. By keeping 
Bell's model ${\bf M}_{\rm{Bell}}$ as representing physical reality, one confronts with treating of complementarity as the fundamental property of (observational) microphenomena. 
     
At the level of correlations, 
\begin{equation}
\label{j1}
\langle A_i B_j\rangle_\rho =\rm{Tr} \hat \rho \hat A_i \hat B_j= 
 \lim_{N\to \infty} \frac{1}{N} \sum_{k=1}^N A_{ik} B_{jk},
\end{equation}
where $(A_{ik}), (B_{jk})$ are observables' outcomes. At the same time, for the probability distribution 
$P_\rho$ such that $\hat \rho =J_S(P_\rho),$ we have
\begin{equation}
\label{j2}  
\langle A_i B_j\rangle_{P_\rho}= \int_\Lambda A_i(\lambda) B_j(\lambda) P_\rho(d\lambda)=
\lim_{N\to \infty} \frac{1}{N} \sum_{k=1}^N A'_{ik} B'_{jk}
\end{equation}
where $A'_{ik}$ and  $B'_{jk}$ are outcomes of random variables $A_i= A_i(\lambda)$ and $B_j= B_j(\lambda).$ However, since these outcomes can be identified with the outcomes of the quantum observables, $A'_{ik}=A_{ik},
 B'_{jk} =B_{jk},$ we can write
\begin{equation}
\label{j3}
\langle A_i B_j\rangle_{P_\rho}= \lim_{N\to \infty} \frac{1}{N} \sum_{k=1}^N A_{ik} B_{jk}.
\end{equation}
But the same reasoning is applicable to the subquantum random variables $A_1=A_1(\lambda), A_2=A_2(\lambda)$ and $B_1=B_1(\lambda), B_2=B_2(\lambda)$ representing 
the incompatible quantum observables:
\begin{equation}
\label{j4}
\langle A_1 A_2\rangle_{P_\rho}= \int_\Lambda A_1(\lambda) A_2(\lambda) P_\rho(d\lambda) =
= \lim_{N\to \infty} \frac{1}{N} \sum_{k=1}^N A'_{1k} A'_{2k},
\end{equation}
\begin{equation}
\label{j5}
\langle B_1 B_2\rangle_{P_\rho}= \int_\Lambda B_1(\lambda) B_2(\lambda) P_\rho(d\lambda) =\lim_{N\to \infty} \frac{1}{N} \sum_{k=1}^N B'_{1k} B'_{2k}.
\end{equation}
Again by identifying the values of subquantum and quantum observables, we obrtain:
\begin{equation}
\label{j6}
\langle A_1 A_2\rangle_{P_\rho}= \lim_{N\to \infty} \frac{1}{N} \sum_{k=1}^N A_{1k} A_{2k},
\end{equation}
\begin{equation}
\label{j7}
\langle B_1 B_2\rangle_{P_\rho}= \lim_{N\to \infty}  \frac{1}{N} \sum_{k=1}^N B_{1k} B_{2k}.
\end{equation}
Representations (\ref{j6}), (\ref{j7}) of subquantum correlations (within CTM ${\bf M}_{\rm{Bell}})$ via 
outcomes of observables supports the assertion that the subquantum correlations 
$\langle A_1 A_2\rangle_{P_\rho}, \langle B_1 B_2\rangle_{P_\rho}$ should be measurable (at least in principle
and in future). It is not clear how Bell would treat this objection to his argument. I guess that he would agree that his model  with hidden variables, ${\bf M}_{\rm{Bell}},$ collides with the complementarity principle. However, he might choose to move between Scylla and Charybdis:
\begin{itemize}
\item {\bf S} identification the values of subquantum random variables with the values of quantum observables;
\item {\bf Ch:} the complementarity principle,     
\end{itemize}
and claim that {\bf S} and {\bf Ch} can peacefully coexist. He might say that (\ref{j3}) is legal and hence the experimental verification of  ${\bf M}_{\rm{Bell}}$ is possible, but (\ref{j6}), (\ref{j7}) are illegal
and treatment of these correlations as experimentally verifiable is forbidden. For me the later position is  inconsistent (although logically possible). This inconsistency was the basis of  De Broglie's critique of Bell's argument \cite{DeBroglie} (section \ref{DBL}).

This is the good place to recall that the physical seed of the complementarity principle is in Bohr's quantum postulate on the existence of indivisible quantum of action given by the Planck constant $h:$ 
incompatible observables exist due to the existence in nature of the minimal action \cite{BR1}-\cite{BR2a} (see \cite{KHR7b}). Thus, Bell's conflict with the complementarity principle is in fact the conflict with the quantum postulate - the existence of $h.$ Hence, this is the conflict with the very foundation of quantum physics, e.g., the quantum model of the black body radiation and processes of spontaneous and stimulated emissions. 

\section{De Broglie's critique of no-go theorems of von Neumann and Bell}
\label{DBL}

Nowadays it is practically forgotten that De Broglie considered all no-go theorems by which hidden variables for QM do not exist as totally misleading \cite{DeBroglie}. His double solution model \cite{DB} can be considered as a model with hidden variables of the field type. The pilot wave is a hidden variable.
\footnote{ In this aspect the double solution model and PCSFT are similar.  Schr\"odinger's attempt to find a subquantum model of the wave type as well as Einstein's search for a subquantum nonlinear field model were also steps in the same direction, although, as was noted, methodologically their positions are different.  Schr\"odinger tried to realize the Bild approach for QM, treated as OM, but Einstein dreamed for a model carrying jointly the features of CTM and OM.} As we shall see, De Broglie followed the Bild-conception  without knowing about it.   

To justify his double solution model and its peaceful coexistence with QM, De Broglie criticized the most famous no-go theorems, the von Neumann and Bell theorems \cite{VN, Bell1}. He did not criticize the mathematical derivation of these theorems, but their interpretation and straightforward identification of 
hidden and observational quantities. His interpretation of these theorems was presented in very detail by Lochak \cite{Lochak,Lochak1,Lochak2}. Paper \cite{Lochak1} is available for free reading via Google books; we cite it: 

\begin{footnotesize}
Von  Neumann proved a theorem which claims that there are no pure states without statistical dispersion. This result is indeed intuitively obvious because the absence of dispersion in a pure state mean that it would be possible to measure simultaneously the physical quantities attached to a system described by this state. But in fact we know that this is impossible for non-commuting quantities. In this sense, the theorem is nothing but a consequence of Heisenberg's uncertainties.
\end{footnotesize}

From this one can conclude that a pure state of QM can't be considered as representing an ensemble of systems following the laws of classical probability theory (see \cite{Lochak1} for detailed discussion)
and quantum observables as classical random variables. From this, von Neumann made the conclusion that generally it is impossible to create any model with hidden variables behind QM. As pointed out in \cite{Lochak1},   
    
\begin{footnotesize} 
De Broglie's answer consists essentially in asserting that if any hidden parameters do exist, they cannot obey quantum mechanics because if you try to imagine hidden parameters it is of course in order to restore the classical scheme of probabilities. Now if you need a classical scheme of probabilities for objective(but hidden) values of physical quantities which are introduced in quantity of hidden parameters,
these probabilities cannot be probabilities observed on the result of a measurement: simply because the 
observed probabilities do obey the quantum scheme and not the classical one!
\end{footnotesize}

Hence, not only hidden parameters $\lambda \in \Lambda,$ but even variables $A=A(\lambda)$ are hidden 
and the probability distributions of these variables $P_A$ should not be identified with quantum probability distributions. 

For De Broglie, it was evident that classical and quantum probability calculi differ crucially, by attempting to apply the former for quantum observables one immediately confronts with the Heisenberg uncertainty principle (and Bohr's complementarity principle\cite{BR0,PL1,PL2}). This is precisely my viewpoint which was presented in article \cite{NL1a} (see section \ref{BellBild}).

Hence, De Broglie's viewpoint on the interrelation of  subquantum and quantum models matches perfectly the CTM-OM approach. In fact, it matches the ontic-epistemic framework, since De Broglie considered hidden variables and physical quantities, $A=A(\lambda),$ as objective entities.  But, as was already noted, schematically CTM-OM  and the ontic-epistemic frameworks are similar. 

De Broglie's statement that quantities of a subquantum theory are hidden and their probability distributions should not be identified with the probability distributions of quantum observables matches the PCSFT-QM coupling considered in section \ref{PCFTBild}. PCFT-quantities are quadratic forms of fields playing the role of hidden variables. Such quantities have the  continuous range of values, but say the quantum spin observables have the discrete spectra. Of course, they cannot have the same probability distribution, even without consideration of correlations. The correspondence between classical probability calculus of PCSFT and the quantum probability calculus is fuzzy, classical covariance  operators  are mapped to density operators, see (\ref{m1q}). 

De Broglie and Lochak used the same argument for the critical analysis of the Bell theorem: one should sharply distinguish subquantum and quantum quantities and not identify their outcomes and probability distributions. Not only hidden parameters are hidden, but also quantities dependent on them and their probability distributions. So, Bell's model is a very special CTM for QM. Yes, it should be rejected, as follows from the quantum formalism and experiments. But its rejection does not prevent search for other CTMs for  QM with more complicated connection between subquantum and quantum quantities and their 
probability distributions. From this viewpoint, the foundational value of the Bell theorem is overestimated. 

I again repeat that it is a pity that the fathers of QM, including De Broglie, were not aware about the works 
Helmholtz, Hertz, and Boltzmann. The Bild-conception  would provide the rigid philosophic basis for establishing proper interrelation between subquantum models with hidden variables and QM.    

\section{Classical mechanics}

For classical mechanics, CTM and OM coincide. On one hand, this was fortunate for development of physics, since it simplified so much its philosophic basis and highlighted the role of observation. On the other hand, identification of one special mathematical model, Newtonian mechanics, with reality supported similar ontic treatment of all physical models. The ontic viewpoint of a scientific theory dominated during a few hundreds years, up to works of Helmholtz, Hertz, Boltzmann, and Schr\"odinger. However, these works did not revolutionized the philosophy of science. For example, acausality of QM is still considered as the property of nature, so to say irreducible quantum randomness is ontic.             

We note that it seems that Hertz didn't consider classical mechanics as OM, see again \cite{Hertz1}:
\begin{footnotesize}
If we try to understand the motions of bodies around us, and to refer them to simple
and clear rules, paying attention only to what can be directly observed, our
attempt will in general fail.  
\end{footnotesize}

This statement definitely refers to classical mechanics. Similarly to Hertz, Atmanspacher \cite{ATM1}
also considered the two level description even for classical physics and suggested the corresponding mathematical examples.

The need of separate OM model for classical mechanics became clear in the process of 
creation of the mathematical description of the Brownian motion which will be considered in the next section and CTM-OM structured in accordance with the Bild-conception ion.   
  
\subsection{Brownian motion: two levels of description  from the time-scale separation}
\label{SBrownian motion}

Here we follow the article of  Allahverdyan, 
Khrennikov, and Nieuwenhuizen \cite{AKNJ}: 

\begin{footnotesize} 
The dynamics of a Brownian particle can be observed at two levels
\cite{risken}. Within the first, more fundamental level the Brownian
particle coupled to a thermal bath at temperature $T$ is described via
definite coordinate $x$ and momentum $p$ and moves under influence of
external potential, friction force, and an external random force.  The
latter two forces are generated by the bath. The second, overdamped
regime applies when the characteristic relaxation time of the
coordinate $\tau_x$ is much larger than that of the momentum $\tau_p$,
$$
\tau_x\gg\tau_p
$$ (overdamped regime). On times much larger than
$\tau_p$ one is interested in the change of the coordinate and defines
the {\it coarse-grained} velocity as $v=\Delta x/\Delta t$ for
$\tau_x\gg \Delta t\gg \tau_p$.  This definition of $v$ is the only
operationally meaningful one for the (effective) velocity within the
overdamped regime.  It appears that the coarse-grained velocity,
though pertaining to single particles, is defined in the context of
the whole systems of coupled Brownian particles.
\end{footnotesize}

The evolution of the momenta of the Brownian particles is very fast and cannot be resolved on the 
time-scales available to the experiment. To obtain experimentally 
accessible quantities, one employ the technique of the time-scale separation and measurement of the coarse grained velocity and osmotic velocity. These quantities can be measured. They are assigned not to an individual Brownian particle, but to an ensemble of particles coupled to bath, so these are statistical quantities.    

In terms of the present article, Brownian motion is described by CTM ${\bf M}_{CTMB}$ with the
 phase space $(x,p)$ and OM ${\bf M}_{\rm{OMB}}$ with the coarse grained velocities $v_+, v_-$ or the osmotic velocity $u=v_+ -v_-,$ 
The later description is based on observational quantities $(x,u).$ As was shown in article \cite{AKNJ},  ${\bf M}_{\rm{OMB}}$ shows some properties of QM, e.g., there are  analogs of the Heisenberg uncertainty relations and entanglement; in particular, for a pair of Brownian particles
the joint probability distribution $P(t, x_1,u_1, x_2, u_2)$ does not exist. Of course, the OMs
${\bf M}_{\rm{OMB}}$ and ${\bf M}_{\rm{QM}}$ differs essentially. For example, for a single particle the probability distribution $P(t,x,u)$ is well defined, incompatibility appears only in compound systems.
  
Nowadays the above two level structuring of scientific theory of the Brownian motion is shaken by the novel 
experimental possibilities for the measurement of momentum $p$ of a Brownian particle. A variety of experiments was performed during the last years (see,e.g., \cite{BrowianV}). In spite of some diversity in experimental outputs, it is clear that experimental science is on the way to establishing the robust procedures for measurement of the Brownian momentum $p$. Through experimental research,  CTM  
${\bf M}_{\rm{CTMB}}$ is getting the OM-status. However, new theoretical efforts are needed to merge  
${\bf M}_{\rm{CTMB}}$ and ${\bf M}_{\rm{OMB}}$ treated as OMs. The osmotic velocity $u$ (an element of 
${\bf M}_{\rm{OMB}})$ is not straightforwardly derived within ${\bf M}_{\rm{CTMB}}.$ At least for me, connection between the velocity and coarse grained velocity is not clear. How is the latter derived from the former?      

This special example supports the search for CTMs for QM (see the discussion at the end of 
section \ref{use}). Some hidden quantities of such models can serve as the candidates for the future experimental verification. One of the problems of such project is that, since creation of QM, physicists (mathematicians and philosophers) created too many subquantum models operating with a variety of hidden quantities, as say the quantum potential in Bohmian mechanics or the random field in PCSFT. What are the most probable candidates for future experimental verification? The Bell hidden variable model 
\cite{Bell0,Bell1} is one of CTMs for QM, it can be directly tested experimentally. It was tested and rejected.  

 \section{Discussion on the Bild-conception  and its role in foundations of science}
 
My aim  is to recall to  physicists and especially to experts in foundations (not only of quantum physics, but also classical mechanics and field theory, statistical mechanics, and thermodynamics) about works of 
Helmholtz, Hertz, and Boltzmann \cite{Helmholtz,Hertz,Hertz1,Boltzmann,Boltzmann1} on the meaning of a scientific theory which led to the Bild-conception  - the mathematical model concept of a scientific theory. By appealing to the two level description of natural phenomena, CTM-OM description it is possible to resolve many foundational problems, including acausality of QM. Moreover, the Bild-conception  demystify quantum foundations. The ``genuine quantum foundational problems'' such as the possibility to introduce hidden variables were discussed long ago. The latter problem was analyzed by Hertz who tried to reduce the classical electromagnetic field to an ensemble of mechanical 
oscillators \cite{Hertz}. From the viewpoint of the Bild-conception , Bell's attempt to invent hidden variables for QM 
is very naive; if such variables were existed their coupling with quantum observables might be not as straightforward as in the Bell model. Within the Bild-conception , it becomes clear why Schr\"odinger did not consider acausality of  quantum observations as the barrier on the way towards a causal description of quantum phenomena \cite{SCHB}-\cite{SCHB1}. It seems that similarly to Bell \cite{Bell1}, von Neumann was neither aware about development of the philosophy of science by the German school of physicists in 19s century. He treated the quantum measurement problem too straightforward and acausality and irreducible quantum randomness appeared as consequences of such treatment \cite{VN}. He did not appeal to the two level CTM-OM description of microphenomena. 

In a series of works \cite{KHRB1}-\cite{KHRB3}, Krennikov et al. developed PCSFT, CTM with classical random fields, reproducing QM interpreted as OM for microphenomena. However, PSCFT-QM coupling is not so simple as in the Bell framework. 

The two level description of physical phenomena is in fact widely used in statistical physics and it sis based on time-scales separation technique and consideration of coarse quantities. All such descriptions are well accommodated within the Bild-conception . The Brownian motion in the overdamped regime is described by OM which is not directly coupled to CTM based on the classical mechanical description. 

Finally, we remark that Primas-Atmanspacher \cite{Primas,ATM} ontic-epistemic approach to physical theories 
(see also, e.g., \cite{ATM1,ATM2} is formally similar to the Bild-conception . But in accordance with the Bild concept, no model describes reality as it is.

\section*{Acknowledgments} 
This work was partially supported by COST EU-network DYNALIFE, Information, Coding, and Biological Function: the Dynamics of Life, CA211 69.

\section*{Appendix:}

Here we follow the article of Allahverdyan, Khrennikov, and Nieuwenhuizen \cite{AKNJ}.

The system under analysis consists of  $N$ identical Brownian particles with coordinates
${\bf x}=(x_1,...,x_N)$ and mass $m;$  particles interact with thermal baths at temperatures 
$T_i$ and coupled via  a potential
$U(x_1,...,x_N)$.  We consider so.called {\it overdamped limit} \cite{risken}:

\begin{itemize}
\item The characteristic relaxation time of particles' momenta $p_i= m\dot{x}_i$ is essentially less
than the characteristic relaxation time of the coordinates: $
\tau_x\gg\tau_p.$ 
\item  Dynamics is considered in the time-range:
 $$
\tau_p << t \leq \tau_x .
$$
\end{itemize}

The conditional probability $P({\bf x},t|{\bf x}',t')$ satisfies the Fokker-Planck equation 
(the special case of the Kolmogorov equation for diffusion):
\cite{risken}:
\begin{equation}
\label{fokkerplanck}
\partial_t P({\bf x},t|{\bf x} ',t')
=-\sum_i\partial_{x_i}\,[\, f_i({\bf x})\,P({\bf x},t|{\bf x} ',t') \,]+
\sum_iT_i\,\partial^2_{x_ix_i}P({\bf x},t|{\bf x} ',t'),\qquad
t\geq t',
\end{equation}
with the initial condition (corresponding to the definition of conditional probability)
\begin{equation}
\label{d2}
P({\bf x},t|{\bf x}',t)= \delta({\bf x}-{\bf x}')
\equiv\prod_{i=1}^N\delta(x_i-x'_i).
\end{equation}

Now consider an ensemble $\Sigma({\bf x},t)$ of all realizations of the 
$N$-particle system having at time $t$ the fixed coordinate vector
${\bf x}$. This ensemble of systems is chosen out of all possible realizations
for measuring particles' coordinates.  For $\Sigma({\bf x},t),$ the average coarse-grained
velocity for the particle with index $j$ might be heuristically defined as 
\begin{equation}
v_{j}({\bf x},t)={\rm lim}_{\epsilon\to 0}
\,\int\d {\bf y}\,\frac{y_j-x_j}{\epsilon}\,P({\bf y},t+\epsilon|{\bf x},t).
\end{equation}
However, irregularity of the Brownian trajectories implies non-existence of this limit; so one should to define  the velocities for different directions of
time \cite{nelson}:
\begin{equation}
\label{dish1}
v_{+,j}({\bf x},t)={\rm lim}_{\epsilon\to +0}
\,\int\d y_j\,\frac{y_j-x_j}{\epsilon}\,P(y_j,t+\epsilon|{\bf x},t),
\end{equation}
\begin{equation}
\label{dish2}
v_{-,j}({\bf x},t)={\rm lim}_{\epsilon\to +0}
\,\int\d y_j\,\frac{x_j-y_j}{\epsilon}\,P(y_j,t-\epsilon|{\bf x},t).
\end{equation}

What is the physical meaning of these expressions?
The directional coarse grained velocity $v_{+,j}({\bf x},t)$ is the average 
velocity to move anywhere starting from $({\bf x},t)$, whereas
$v_{-,j}({\bf x},t)$ is the average velocity to come from anywhere and to
arrive at ${\bf x}$ at the moment $t$.

For the overdamped Brownian motion, almost
all trajectories are not smooth and this is the reason for 
\begin{equation}
\label{bab1}
v_{+,j}({\bf x},t)\not=v_{-,j}({\bf x},t).
\end{equation}
The difference 
\begin{equation}
\label{bab2}
u({\bf x},t)=v_{+,j}({\bf x},t)- v_{-,j}({\bf x},t)
\end{equation}
characterizes the degree of non-smoothness; it is called osmotic velocity; analitically it can be represented in the form
\begin{equation}
\label{babek}
u_j({\bf x},t)=\frac{v_{-,j}({\bf x},t)
-v_{+,j}({\bf x},t)}{2}=-T_j\partial_{x_j}\ln P({\bf x},t).
\end{equation}

If we consider $\epsilon$ much smaller than the
characteristic relaxation time of the momentum (
apply definitions (\ref{dish1}) and (\ref{dish2}) to a smoother
trajectory), then $v_{+,j}({\bf x},t)$ and $v_{-,j}({\bf x},t)$ will be equal
to each other and equal to the average momentum.

\end{document}